\documentclass[a4paper]{article}

\usepackage{INTERSPEECH2022}
\usepackage{mathtools,amsmath,graphicx,array}
\usepackage{amsthm,amsmath,amssymb}
\usepackage{mathrsfs}
\usepackage{comment}
\usepackage{subfigure}
\usepackage{longtable}
\usepackage{setspace}
\usepackage{booktabs}
\usepackage{multirow}
\usepackage{algorithm}
\usepackage{algpseudocode}
\usepackage{float}
\usepackage{graphicx}
\usepackage[colorlinks,linkcolor=green]{hyperref}

\title{Adversarial Privacy Protection on Speech Enhancement}
\name{Mingyu Dong, Diqun Yan, Rangding Wang}
\address{
  College of Information Science and Engineering, Ningbo University
  }
\email{1253173217@qq.com, yandiqun@nbu.edu.cn, wangrangding@nbu.edu.cn}

\begin{document}

\maketitle
\begin{abstract}
  Speech is easily leaked imperceptibly, such as being recorded by mobile phones in different situations. Private content in speech may be maliciously extracted through speech enhancement technology. Speech enhancement technology has developed rapidly along with deep neural networks (DNNs), but adversarial examples can cause DNNs to fail. In this work, we propose an adversarial method to degrade speech enhancement systems. Experimental results show that generated adversarial examples can erase most content information in original examples or replace it with target speech content through speech enhancement. The word error rate ($WER$) between an enhanced original example and enhanced adversarial example recognition result can reach 89.0\%. $WER$ of target attack between enhanced adversarial example and target example is low to 33.75\% . Adversarial perturbation can bring the rate of change to the original example to more than 1.4430. This work can prevent the malicious extraction of speech. Code is available in \href{https://github.com/DanMerry/Advsarial_SE}{Github}
\end{abstract}

\section{Introduction}
\label{sec:intr}

	Private content information is easily extracted from leaked speech, which may be recorded by some mobile phone applications. The extraction process   depends on many speech analysis tasks. Speech enhancement is a common technology used in many tasks, such as speech recognition  \cite{nassif2019speech} and voice conversion  \cite{jia2018transfer}. For example, the enhanced noisy speech can improve the performence of a speech recognition system \cite{weninger2015speech}. Speech enhancement models have developed rapidly along with   deep neural networks (DNNs). The adversarial example is the deadly weakness of DNN models which is a technology that can cause a DNN model to output a wrong result, and the generated adversarial example is imperceptible to humans. In this work we proposed an adversarial attack method to prevent our leaked speech from maliciously extracting.

	The application scenario of the adversarial example on speech enhancement can be described as follows. Our speech may be monitored if we carry with mobile device, and maliciously recorded speech will be sent to extract useful content information. The usual method for protecting leaked speech  is to add specific adversarial perturbation which can cause speech recognition systems misrecognize\cite{pmlr-v97-qin19a}. However, the speech may be enhanced before recognition, and the speech enhancement systems will erase the useful adversarial perturbation. To solve this problem, we propose an adversarial example to attack speech enhancement models directly, which causes the enhanced example to be recognized as a target phase or as meaningless words. As shown in Figure 1, the recorded speech through the adversarial protect method will be denoised to one useless speech which is hard to extract some useful private content information. And in mobile processor, the exreacted useless content information will be discarded. In this way, the privacy in speech is protected.

	\begin{figure}[!t]
		\centering
		\includegraphics[width=0.50\textwidth]{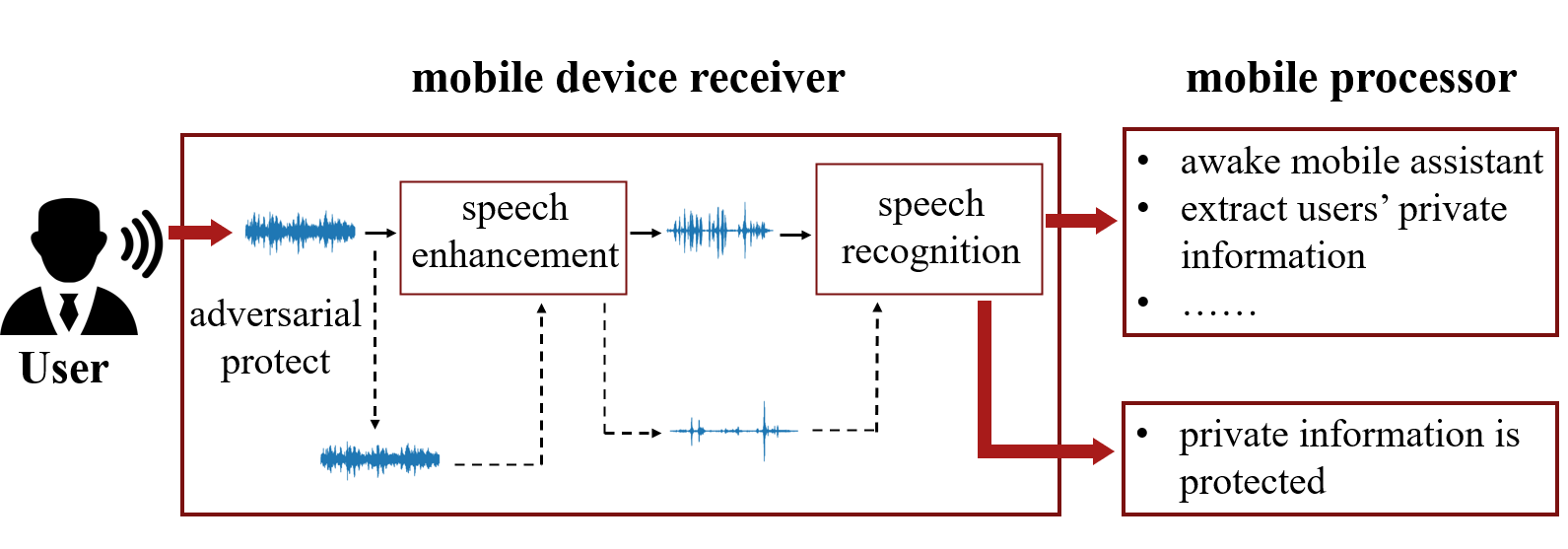}
		\caption{The process of privacy protect in mobile phone terminate}
		\label{2}
	\end{figure}

	Some classical methods of adversarial examples have developed since being proposed in 2013   \cite{szegedy2013intriguing}. Adversarial examples have always worked on the image domain in recent research, with the image classification task as the target model. As attention became more focused on the audio domain, research on audio adversarial examples grew. Kos \cite{kos2018adversarial} proposed adversarial example to attack the image reconstruction model, which is the first work related to the deep generation model. In the audio domain, Takahashi proposed an adversarial example to attack an audio source separation model that can help protect songs' copyrights from the abuse of separated signals  \cite{9414844}. Chien-yu Huang proposed  work of attacks on voice conversion systems  \cite{huang2021defending}, which can protect a speaker's private information.

	However, adversarial examples have some deficiencies: (a) The choice of target models is limited, and research on the classification task accounts for majority. Thus, research on the generation task still needs to be done; and (b) the application scenario of adversarial example is restricted, and is mainly used to enhance the model's robustness. The various usages in adversarial examples should be expanded, such as to protect privacy.

	The contributions of this paper are as follows:
	\begin{itemize}
		\item We believe this is the first adversarial example work on the speech enhancement systems. We combine practical scenes to generate adversarial examples with different $SNRs$, and propose new evaluation metrics;
		
		\item We analyze the transferability between speech enhancement models, using different targets to make the enhanced adversarial example show different effects;
		
		\item From the application scene, the adversarial example in our work is expanded to protect the privacy of leaked speech.
	\end{itemize}
	
	The rest of this paper is organized as follows. Section II   introduces some work related to the proposed method, which is described in section III. Section IV discusses experiments. Section V provides our conclusions.

\section{Related work}

	\subsection{Gradient-based adversarial method}

	The gradient-based method is efficient, and it provides the primary idea  for the adversarial example. The classical fast gradient sign method (FGSM)  \cite{goodfellow2014explaining} is a shortcut algorithm that simply adds a signed gradient to the original example:
	\begin{equation}
		x^{*} = x + \epsilon*sign(\nabla_{x} \mathcal{L}(f(x), t_{adv})),
	\end{equation}
	where $\epsilon$ controls the steps of perturbation, and $\mathcal{L}(\cdot)$ is the loss between the predicted result and the target $t_{adv}$. It is a one-step, white-box method that can be quickly applied in most open-source models.

	Projected gradient descent (PGD)  \cite{madry2017towards} is an iterative method to generate a more effective adversarial example based on FGSM. Compared to the one-step attack, the generated perturbation by PGD is smaller in every step, in which the value of the perturbation will be small and limited to a specific range:
	\begin{equation}
		x_{t+1}^{*} = \prod_{x+s}( x_{t}^{*} + \epsilon*sign(\nabla_{x} \mathcal{L}(f(x_{t}), t_{adv}))),
	\end{equation}
	where $s$ is the range of adversarial perturbation.   PGD has the advantage that  the generated adversarial examples have a higher attack success rate, but the adversarial perturbation in many steps will be magnified, so the attacked example will be greatly changed.

	\subsection{Optimization-based adversarial method}
	
	The Carlini and Weanger  (C\&W) method \cite{carlini2017towards} is a classical optimization-based method, whose core concept  is to optimize an example through the loss function to measure the perceptibility of the adversarial example and the output results. The structure and parameters of the target model are not a necessary condition. If we set the original example as the optimization target, the gradient in model is not needed and it can be seen as a black-box method. Its advantage is that the generated adversarial example is much more imperceptible, but the optimization process requires time.

	\subsection{Speech enhancement system}
	Speech enhancement systems are the target of adversarial examples, which can tranform noisy speech to clear speech. Recently,  DNN-based speech enhancement technology has developed quickly. So, we choose the advanced MetricGAN+ \cite{fu2021metricgan+} and classical SEGAN \cite{pascual2017segan} as our attack targets. MetricGAN+ is based on MetricGAN\cite{fu2019metricgan}, and its perceptual evaluation of speech quality (PESQ) can reach 3.15, which is a state-of-the-art result. SEGAN\footnote{$https://github.com/santi-pdp/segan\_pytorch$}  applies U-Net in speech enhancement, and its PESQ can reach 2.16.

\section{proposed methods}

	In this section, we expand the adversarial attack method from classification task to generation task. To choose an appropriate target is an important part in generating an adversarial example, as different targets can exhibit different effects. A target can be a silent audio  clip or random noise. Targets have different aims. Experimental results show that setting a  silent clip as a target can cause the predicted results to lack content information, and to set random noise as a target can cause the predicted results to sound chaotic. To highlight the effect of an adversarial attack, we   choose a silent audio  clip as a target  in the following experiments. In speech enhancement models, the enhancement process can be described as 
	\begin{equation}
		y_{en} = g(x_{noisy}),
	\end{equation}
	where $g()$ is the speech enhancement process that can transform  noisy speech  to clear speech, $x_{noisy}$ is a noisy example, and  $y_{en}$ is an enhanced example.
	
	There is the issue of how to make gradient-based methods work on a speech enhancement model. Our adversarial methods are based on the above gradient-based methods of FGSM and PGD. Most existing methods are designed to attack the classification task, hence the cross-entropy loss function in the original methods is not suitable for the generation task.  We use the mean squared error (MSE) and norm as our loss function. The loss between the predicted  and target results in the one-step FGSM and iterative PGD are replaced by the MSE. The usage of the MSE loss function in FGSM and PGD can be described as follows: 
	
	\begin{equation}
		x^{*} = x + \epsilon*sign(\nabla_{x} MSE(g(x), t_{adv})),
	\end{equation}

	\begin{equation}
		x_{t+1}^{*} = \prod_{x+s}( x_{t}^{*} + \epsilon*sign(\nabla_{x} MSE(g(x_{t}), t_{adv}))),
	\end{equation}
	
	Finally, there is the issue of how to make an optimization-based (OPT) method work on a speech enhancement model. We set the original example as our optimization target. The loss function includes the distance between the original and  adversarial example and distance between the predicted result and the target. The proposed method can be described by Algorithm 1.
	
	\renewcommand{\algorithmicrequire}{\textbf{Input:}} 
	\renewcommand{\algorithmicensure}{\textbf{Output:}}
	\begin{algorithm}[H]
		\caption{Optimization-based method}
		\label{alg:1}
		\begin{algorithmic}[H]
			\Require
			Original example $x_{noisy}$, target $t_{adv}$, iterations $itr$
			\Ensure
			Targeted adversarial example $x^{*}_{noisy}$ 
			\State $x^{*}_{noisy}$ $\leftarrow$  $x_{noisy}$
			\For{$itr$ step}
			\State $\mathscr{L}_{1}$ $\leftarrow$ $MSE(x^{*}_{noisy}, x_{noisy})$
			\State $\mathscr{L}_{2}$ $\leftarrow$ $\left \|g(x^{*}_{noisy})-t_{adv}\right\|_{2}$
			\State $\mathscr{L}_{3}$ $\leftarrow$ $\mathscr{L}_{1}$ + $\alpha$ * $\mathscr{L}_{2}$
			\State Update $x^{*}_{noisy}$ by Adam optimization
			\EndFor
			\State return $x^{*}_{noisy}$

		\end{algorithmic}
	\end{algorithm}

\section{Experiments}

	\subsection{Experimental setup}
	
	Voicebank \cite{veaux2013voice} and Demand \cite{thiemann2013demand} are chosen as the datasets of clean speech and background noise, respectively. Voicebank contains more than 500 hours of recordings from about 500 healthy speakers, and Demand contains 6 indoor noise audios and 12 outdoor noise audios. We used   the open-source pretrained model in SpeechBrain  \cite{ravanelli2021speechbrain} as the speech recognition system. We provide the experimental results for different $SNR$ values which include -8, -4, 0, 4 and 8 on five different noisy scenes which include busy subway station, public town square, public transit bus, private passenger vehicle and subway. The smaller the $SNR$,  the noisier the speech. In OPT method, we set $\alpha$ as 0.0001 which make two loss fuctions in the same weight.

	\begin{table*}[htbp]
		\caption{Recognition results of target attack }
		\renewcommand\tabcolsep{15.0pt} 
		\centering
		\begin{tabular}{ll}
			\toprule
			Item & Recognition results   \\ \midrule
			original ground-truth    & he claimed his insurance company contested the damages not the restaurant  \\
			target ground-truth    & we have to look at everything before we make any final decision
			\\ 
			original enhanced reco.     & he claimed his insurance company and tested the damages not the restaurant  \\
			target enhanced  reco.  & we have to look at everything before we make any final decision \\ 
			adv. enhanced reco.  & we have to look at everything before we make any final decision  \\ \bottomrule
			% $WER$   & 0\% \\ 
		\end{tabular}
	\end{table*} 

\subsection{Evaluation metrics}

The attack effects of adversarial examples on speech enhancement systems are measured differently from those in classification tasks. Two factors are considered: (a) How much does the adversarial perturbation affect  speech enhancement? (b) How bad does the enhanced adversarial speech sound? We use the degree of enhancement $DE$ and residual perturbation rate $RPR$ to measure the effect of adversarial examples. We define $x$, $y$, $x^{*}_{en}$, and $y^{*}_{en}$ as the  input speech, enhanced original speech, adversarial example, and enhanced adversarial example, and
\begin{equation}
	RPR = \frac{\ln\left \|y^{*}_{en}- y_{en}\right \|_{2}}{\ln\left \|x^{*}_{noisy}- x_{noisy}\right \|_{2}},
\end{equation}
\begin{equation}
	DE=F(y^{*}_{en}, gt)-F(y_{en}, gt),
\end{equation}
where $F(x)$   is the word error rate (WER) of speech recognition, and $gt$ is the ground-truth sentence. $DE$ measures how much   the enhanced adversarial example has changed from the original enhanced example. If the adversarial example works, the recognition system will not output a correct result, so there will be a difference between $F(y^{*}_{en}, gt)$ and $F(y_{en}, gt)$. $RPR$ indicates how much the adversarial example affects the enhancement system. In $RPR$, the perturbation is the denominator, and the difference between enhanced original example and enhanced adversarial example is the numerator. So, if $RPR$ is greater than one, then the higher the $RPR$,   the better the degradation effect on the speech enhancement system, meaning that the method works.

	\subsection{Results}
	
		\begin{figure}[htbp]
		\centering
		
		\subfigure[original exmaple]{
			\begin{minipage}[t]{0.5\linewidth}
				\centering
				\includegraphics[width=1.5in]{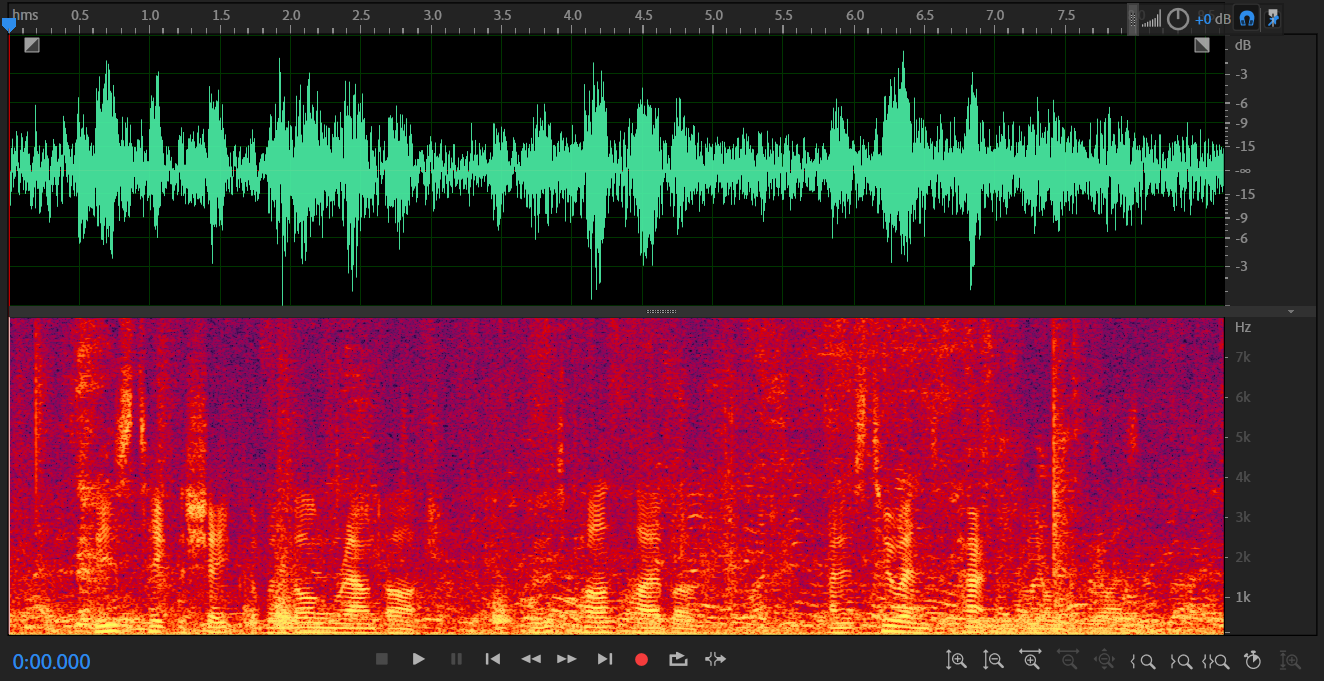}
				
				%\caption{fig1}
			\end{minipage}%
		}%
		\subfigure[adversarial example]{
			\begin{minipage}[t]{0.5\linewidth}
				\centering
				\includegraphics[width=1.5in]{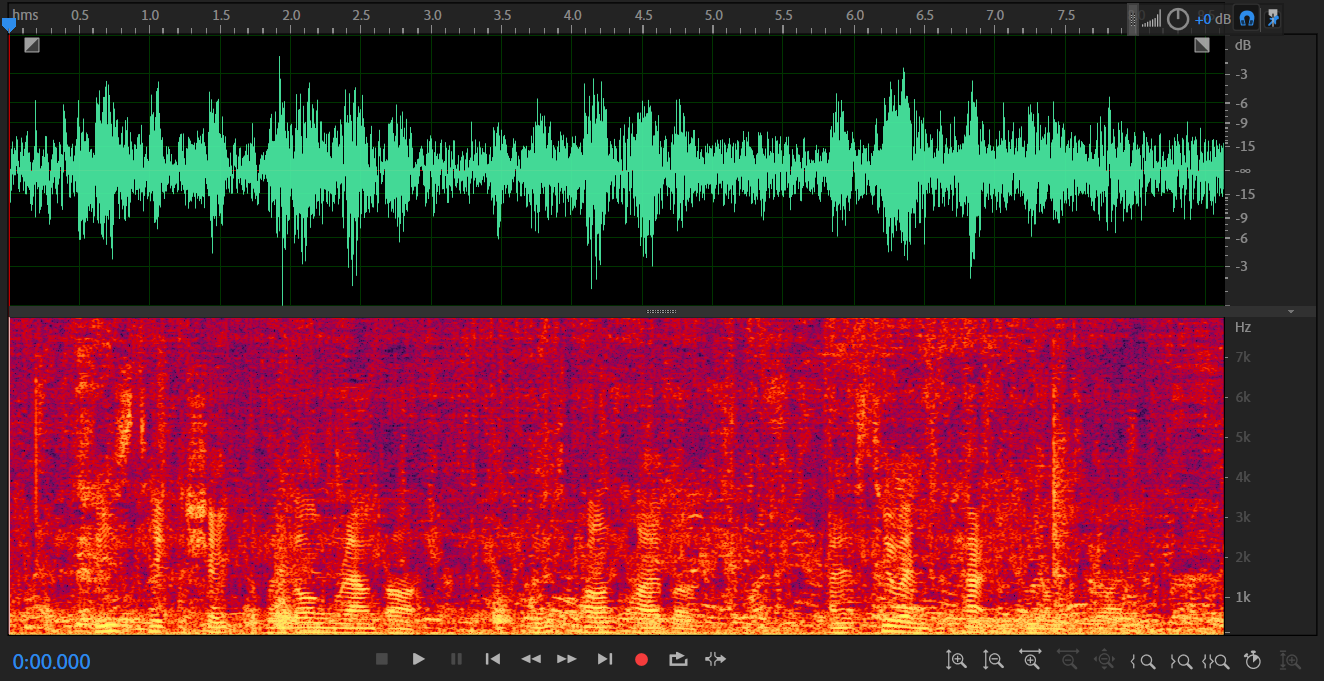}
				%\caption{fig2}
			\end{minipage}%
		}%

		\subfigure[enhanced original example]{
			\begin{minipage}[t]{0.5\linewidth}
				\centering
				\includegraphics[width=1.5in]{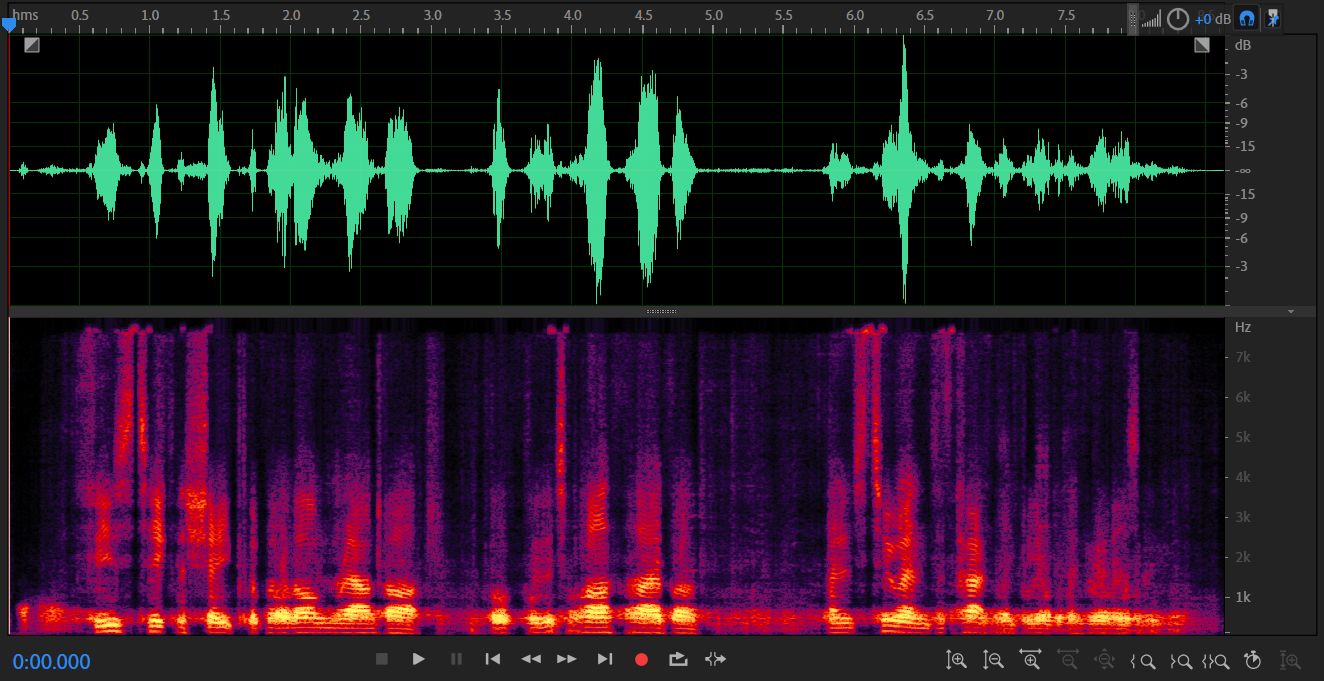}
				%\caption{fig2}
			\end{minipage}
		}%
		\subfigure[enhanced adversarial example]{
			\begin{minipage}[t]{0.5\linewidth}
				\centering
				\includegraphics[width=1.5in]{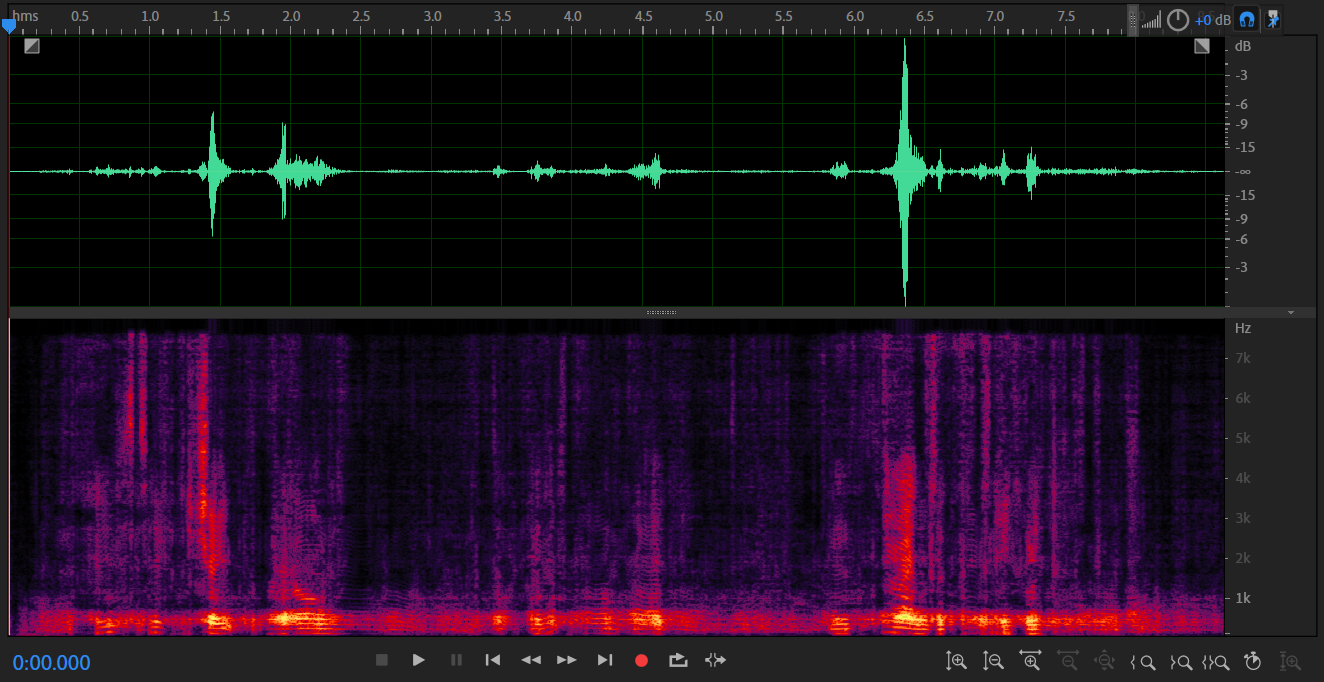}
				%\caption{fig2}
			\end{minipage}
		}%
		
		\centering
		\caption{Effect of adversarial attack works on speech enhancement model}
	\end{figure}
	
	The effect of the adversarial protection can be seen in Figure .2. The generated adversarial exampel is almost imperceptible, but it brings much more changes than perturbation. Much content information in the enhanced adversarial example is erased.

	\subsubsection{Privacy protection effect}
	Figure 3 shows the privacy protection effect on speech enhancement processing. In Figure 3, $WER$ is the averaged value and the adversarial examples are all enhanced. It can be concluded from the figure that there is a difference between the original examples and enhanced examples. $WER$ of speech decreases with the decrease of $SNR$, so it is necessary to denoise the speech before recognition processing.

	\begin{figure}[htbp]
		\centering
		\includegraphics[width=0.50\textwidth]{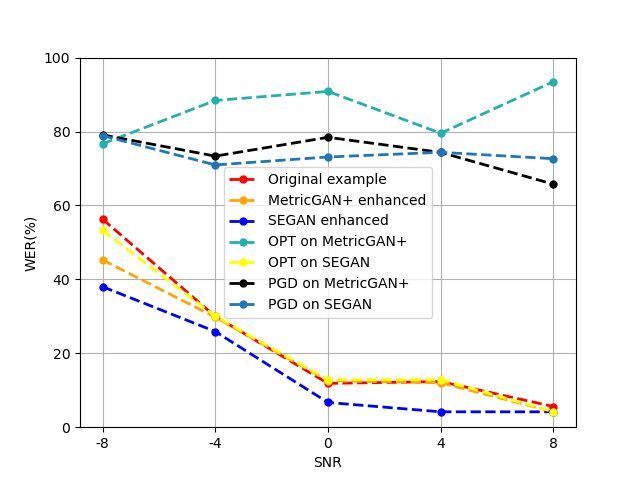}
		\caption{WER results on different SNR examples}
		\label{2}
	\end{figure}

	The performance of $WER$ on enhanced adversarial examples shows that the proposed methods work, and no matter  the value of $SNR$ is, the $WER$ of enhanced adversarial examples maintains a stable level, which means the adversarial examples totally invalidate the speech enhancement system. Due to the SEGAN is a waveform to waveform system, the OPT on the SEGAN converge loss quickly, which make the effect poor. In practical scene, the maliciously recorded speech with some personal content information preprocessed by the proposed method will fail to extract the content information.

	The performance at degrading on target model is shown in Figure 4. The choice of the target model is MetricGAN+ in this experiment. The results show that under different $SNRs$, the OPT method can produce more degradation. Due to the iterative process, OPT can fine-tune a more specific direction to get close to the target in every single step through the loss function. However, FGSM is a one-step method, so the adversarial perturbation may have a rough direction. We can see from Figure 4 that $RPR$ on OPT is higher than FGSM, which means the OPT method can generate a greater difference than FGSM. And as $SNR$ increases, $RPR$ increases. However, $DE$ decreases as $RPR$ increases. The original enhanced examples have a high $WER$, so the difference between the $WER$ values becomes small.

	\begin{comment}
	\begin{table}[htbp]
	\caption{Experimental Results}
	\renewcommand\tabcolsep{15.0pt} 
	\centering
	\begin{tabular}{llll}
	\toprule
	Method                & SNR & RPR    & DE      \\ \midrule
	\multirow{5}{*}{FGSM} & -8  & 0.3286 & 6.9159  \\
	& -4  & 0.4769 & 3.0876  \\
	& 0   & 0.5311 & 11.9619 \\
	& 4   & 0.6159 & 2.1918  \\
	& 8   & 0.5900 & 0.8219  \\ \hline
	\multirow{5}{*}{OPT}  & -8  & 1.6058 & 38.6485 \\ 
	& -4  & 1.4430 & 62.5185 \\
	& 0   & 1.3036 & 84.1638 \\
	& 4   & 1.2519 & 75.3741 \\
	& 8   & 1.2541 & 89.3485 \\ \bottomrule
	\end{tabular}
	\end{table}
	\end{comment}

	\begin{figure}[htbp]
		\centering
		\includegraphics[width=0.50\textwidth]{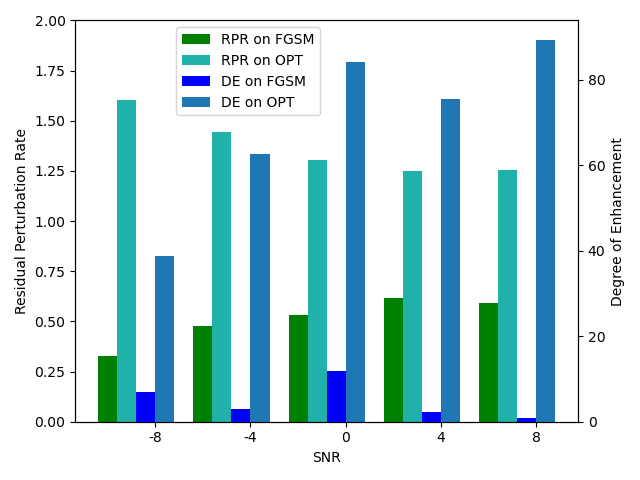}
		\caption{Performance results of   adversarial examples on proposed evaluation metrics}
		\label{2}
	\end{figure}

	\subsubsection{Target attack on speech enhancement}
	To issue an alarm  when the  behavior of malicious information extraction occurs, the target attack on the speech enhancement system can replace the speech content with a warning. The target attack on the speech enhancemnet can achieve the effect of the replacement. In experiments, we choose MetricGAN+ as our target model, with one attacked example and one target example. The chosen attack method was OPT. The $SNR$ of the original example was 0, and the speaker of the target example is different from that of the original example. The aim of the target attack is to make the enhanced example recognize the target's content.
	
	The experimental results of one example are shown in Table 1. We can conclude from the given results that the target attack can completely replace the victim's speech content information. The adversarial enhanced recognition (adv. enhanced reco.) is similar to the target ground-truth. The average $WER$ of enhanced adversarial examples recognition between ground-truth under 6 different $SNR$ can be low to 33.75\%.

	\subsubsection{Transferability between different speech enhancement models}
	Finally, we consider the transferability between different models. In this scenario, the recorded speech is useful for one friendly organization, in which the adversarial examples should not  attack the friendly speech enhancement model, while other models should be successfully attacked by adversarial examples. Speech is leaked easily, and stopping the behavior of malicious extracting from leaked speech is unrealistic. Thus, how to protect the leaked personal information from original source is executable. The transferability of adversarial examples is a significant factor that must be considered. In this experiment, we generate adversarial examples on source model and the PGD is chosen as the attack method. The the generated adversarial examples are tested in the target model.

	\begin{table}[H]
		\caption{Transferability results of adversarial examples between different models}
		\renewcommand\tabcolsep{5.0pt} 
		\centering
		\begin{tabular}{llll}
			\toprule
			source &          target & $DE$ &$RPR$      \\ \midrule
			MetricGAN+  & MetricGAN+ & 58.4087 & 1.2192 \\
			MetricGAN+  & SEGAN & 19.5005 & 0.8575 \\ 
			SEGAN  &      SEGAN & 53.2389 & 1.0758 \\
			SEGAN  &  MetricGAN+ & 41.8050 & 1.3470 \\\bottomrule
		\end{tabular}
	\end{table}

	The results of transferability on adversarial examples can be seen in Table 2. Judging from the $RPR$, the adversarial examples generated by SEGAN can also produce damages to the MetricGAN+, but the $DE$ reduce a little. And the adversarial examples generated by MetricGAN+ are almost invalid to the SEGAN. Due to the SEGAN is a waveform to waveform system, the perturbation is added to the wave directly, the transferability of SEGAN is better than MetricGAN+. It can be concluded that the transferability between different speech enhancement models is weak. This means that an adversarial example can only attack one target model, but will not fail other models. Weak transferability can be used to protect our speech from being destroyed, friendly models are allowed to extract the information in speech, and the target model cannot use this speech.

\section{Conclusion}

In this paper, we proposed an adversarial attck method that works on a speech enhancement system to prevent the malicious extraction of leaked speech without destroying the original examples. Through the speech enhancemnet process , one adversarial example can be enhanced as one silent clip in which the private information is erased rather than one clean speech. Experimental results showed that the adversarial examples can result in  $WER$ values greater than 89\%, and the adversarial perturbation is minute. The target attack can make one enhanced adversarial example recognize as one target phase, the average $WER$ can be low to 33.75\%. We also considered the transferability between different speech enhancement models. This work expand the usage of adversarial example to protect privacy.

In the future work, we intend to improve the transferability to more models. The mainstream speech enhancemnet models are unknown to us, the adversarial examples should attack the applied wildly models successfully which is a totall black-box situation. To accomplish the requirements of transferability, we design to set only one protected speech enhancement model and several victim models.

\section{Acknowledgements}

	This work was supported by the National Natural Science Foundation of China (Grant No. 61300055), Zhejiang Natural Science Foundation (Grant No. LY20F020010), Ningbo Natural Science Foundation (Grant No. 202003N4089).

\bibliographystyle{IEEEtran}

\bibliography{import}

% \begin{thebibliography}{9}
% \bibitem[1]{Davis80-COP}
%   S.\ B.\ Davis and P.\ Mermelstein,
%   ``Comparison of parametric representation for monosyllabic word recognition in continuously spoken sentences,''
%   \textit{IEEE Transactions on Acoustics, Speech and Signal Processing}, vol.~28, no.~4, pp.~357--366, 1980.
% \bibitem[2]{Rabiner89-ATO}
%   L.\ R.\ Rabiner,
%   ``A tutorial on hidden Markov models and selected applications in speech recognition,''
%   \textit{Proceedings of the IEEE}, vol.~77, no.~2, pp.~257-286, 1989.
% \bibitem[3]{Hastie09-TEO}
%   T.\ Hastie, R.\ Tibshirani, and J.\ Friedman,
%   \textit{The Elements of Statistical Learning -- Data Mining, Inference, and Prediction}.
%   New York: Springer, 2009.
% \bibitem[4]{YourName17-XXX}
%   F.\ Lastname1, F.\ Lastname2, and F.\ Lastname3,
%   ``Title of your INTERSPEECH 2022 publication,''
%   in \textit{Interspeech 2022 -- 23\textsuperscript{rd} Annual Conference of the International Speech Communication Association, September 18-22, Incheon, Korea, Proceedings, Proceedings}, 2022, pp.~100--104.
% \end{thebibliography}

\end{document}